\documentclass[prl,superscriptaddress,twocolumn,showpacs,preprintnumbers,amsmath,amssymb]{revtex4}

\usepackage{graphicx}
\usepackage{dcolumn}
\usepackage{bm}

\begin{document}

\preprint{APS/123-QED}

\title{Dynamics of trion formation in GaAs quantum wells}
\author{M.T.~Portella-Oberli}
 \affiliation{Institut de Photonique et Electronique Quantiques,
 Ecole Polytechnique F\'{e}d\'{e}rale de Lausanne (EPFL) CH1015
 Lausanne, Switzerland}
\author{J.~Berney}
 \affiliation{Institut de Photonique et Electronique Quantiques,
 Ecole Polytechnique F\'{e}d\'{e}rale de Lausanne (EPFL) CH1015
 Lausanne, Switzerland}
\author{L.~Kappei}
 \affiliation{Institut de Photonique et Electronique Quantiques,
 Ecole Polytechnique F\'{e}d\'{e}rale de Lausanne (EPFL) CH1015
 Lausanne, Switzerland}
\author{F.~Morier-Genoud}
 \affiliation{Institut de Photonique et Electronique Quantiques,
 Ecole Polytechnique F\'{e}d\'{e}rale de Lausanne (EPFL) CH1015
 Lausanne, Switzerland}
 \author{J.~Szczytko}
 \affiliation{Institut de Photonique et Electronique Quantiques,
 Ecole Polytechnique F\'{e}d\'{e}rale de Lausanne (EPFL) CH1015
 Lausanne, Switzerland}
 \affiliation{Institute of Experimental
 Physics, Warsaw University, Ho\.{z}a 69, 00-681 Warsaw, Poland}
\author{B.~Deveaud}
 \affiliation{Institut de Photonique et Electronique Quantiques,
 Ecole Polytechnique F\'{e}d\'{e}rale de Lausanne (EPFL) CH1015
 Lausanne, Switzerland}

\date{\today}

\begin{abstract}
We propose a double channel mechanism for the formation of charged excitons (trions); they are formed through bi- and tri-molecular processes. This directly implies that both negatively and positively charged excitons coexist in a quantum well, even in the absence of excess carriers. The model is applied to a time-resolved photoluminescence experiment performed on a very high quality InGaAs quantum well sample, in which the photoluminescence contributions at the energy of the trion, exciton and at the band edge can be clearly separated and traced over a broad range of times and densities. The unresolved discrepancy between the theoretical and experimental radiative decay time of the exciton in a doped semiconductor is explained.
\end{abstract}


\maketitle

Positively and negatively charged excitons ($\textrm{X}^+$ and $\textrm{X}^-$ trions) \cite{Kheng1993,Finkelstein1995} are usually compared to their atomic counterparts $\textrm{He}^+$ and $\textrm{H}^-$ respectively. The dynamics of the formation of these atomic ions is of great importance in astronomy~\cite{Frolov2003};  indeed $\textrm{H}^-$ is the primary source of the continuum opacity in most stellar photospheres and contributes to the production of hydrogen and other elements in various parts of the universe. Additionally the abundance of free electrons in the solar atmosphere is indirectly measured in terms of $\textrm{H}^-$ concentration. Lately, models describing the formation dynamics of $\textrm{He}^+$ and $\textrm{H}^-$ have grown more sophisticated and take into account many-body effects resulting from Coulomb correlations and Pauli exclusion principle in partially or fully ionized plasma~\cite{Bi2000}. In semiconductor quantum wells (QWs), trions show a number of properties very similar to excitons~\cite{Chemla1984}, as strong coupling in microcavities~ \cite{Rapaport2001}, absorption bleaching~\cite{Portella2004}, transport~ \cite{Sanvitto2001} and diffusion~\cite{Portella2002} properties, radiative recombination efficiency \cite{Esser2000,Ciulin2000}, and thus have attracted considerable interest. They are highly correlated with the excitons and the plasma of free carriers and offer the possibility to test a model of formation of three particle complexes in this limit. Moreover, trions are promised to play a key role in future applications, notably in quantum-information science \cite{Nielsen2000} and in the future development of all-spin-based scalable quantum computers \cite{Piermarocchi2002,Pazy2003}. In this sense, it is crucial to understand their formation mechanism.

The formation process of neutral excitons ($X$) in QWs has been extensively investigated over the past two decades \cite{Piermarocchi1997,Axt2001} and recently shown to be strongly density and temperature dependent \cite{Szczytko2004}; it is a bi-molecular process, in which an electron ($e$) and a hole ($h$) are bound by Coulomb interaction. Conversely, the formation process of trions has been much less studied. It is largely believed that trions can only be formed if a population of excess carriers is trapped in the well, producing exclusively trions with the same charge. Consequently existing models discriminate the formation channel yielding trions of opposite charge. Such an approach is questionable, especially at low excess carrier densities, where experiments performed on very pure samples demonstrated that both negatively and positively charged excitons do coexist indeed \cite{twotrions}. Current models for trion formation \cite{Jeukens2002,Vanelle2000} surmise that trions are exclusively formed through a bi-molecular process, i.e. the coalescence of an exciton and a charged free carrier. While this is conceivable at low densities, nothing attests that genuine formation of the trion from an unbound electron-hole plasma (tri-molecular formation) is negligible at higher densities.

In this letter, we address this fundamental problem and propose a formation model that fully implements bi- and tri-molecular channels for both negatively and positively charged excitons. We investigate the trion binding dynamics by following separately the evolution of the exciton, trion and plasma luminescence, which is possible by using a time-resolved photoluminescence setup of increased sensitivity. We evidence the complexity of many body effects in the trion formation and demonstrate that all the assumptions made in our model are necessary to adequately describe experiments over a broad range of excess carrier densities. Moreover, we show that the higher the carrier concentration, the more important the tri-molecular process. Theoretical calculations corroborate these result and evidence that momentum conservation of carriers is important in the trion formation process.
\begin{figure}
  \includegraphics[width=\columnwidth]{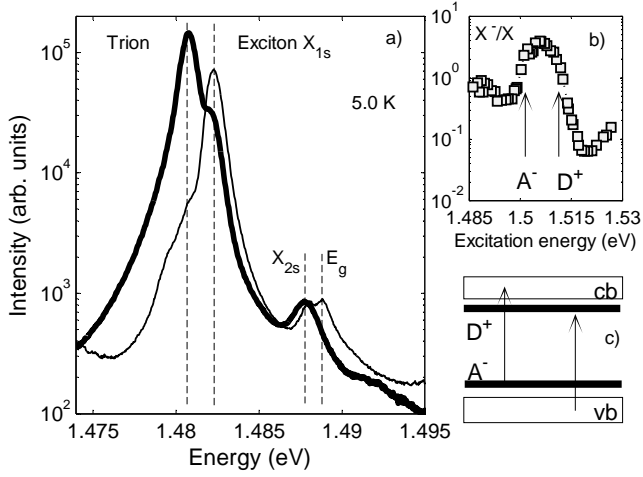}
\caption{\label{fig1} a) CW-luminescence collected for two different
excitation energies: $\hbar\omega$=1.5072 eV (thick line) and
$\hbar\omega$=1.5174 eV (thin line). The structures at 1.4807~eV, 1.4823 (1.4882~eV) and $E_g=1.4888$~eV 
correspond respectively to the trion, heavy-hole exciton $1s$ ($2s$)  and 
plasma transitions (all denoted by vertical dashed lines). b) The
intensity ratio of the trion to exciton transitions as a function of
the excitation energy. c) Schematic diagram of the electronic
transitions from the ionized acceptors $A^{-}$ to the conduction band $cb$
and from the vallence band $vb$ to the ionized donors $D^{+}$.}
\end{figure}

The sample used for this study is a single In$_{x}$Ga$_{1-x}$As 80
\AA\ QW, with an indium content of about $x=5\%$
grown by molecular-beam epitaxy. It was kept at 5~K. The cw photoluminescence spectra were recorded with a CCD camera and time-resolved spectra with a streak camera in photo-counting mode. The temporal resolution of the whole setup was limited to about 10-20~ps, due to the dispersion of the grating, allowing a 0.1~meV spectral resolution. More details about the characterization of the sample and the experimental setup can be found in
\cite{Szczytko2004,Szczytko2004pss}. With proper excitation energy $\hbar\omega$,
we control the electron density accumulated in the well, as shown
in Fig.~\ref{fig1}, where we compare the cw luminescence of the
sample collected for 
$\hbar\omega$=1.5072~eV and $\hbar\omega$=1.5174~eV. Since the trion luminescence is affected by the concentration of charged carriers in the well, the relative $X^{-}/X$ intensity changes with the excitation energy (Fig.~\ref{fig1}b). The carriers trapped into the QW come from the impurities non-intentionally introduced in the GaAs barriers during the growth process, most probably carbon and silicon. Electrons appear when the excitation energy exceeds the energy between ionized acceptors and the conduction band. The excited electrons in the conduction band may then be trapped into the QW. With increasing excitation energy one can transfer electrons from the valence band to ionized donors, thereby increasing the density of holes which then eliminate
electrons (Fig.~\ref{fig1}c). The excess carriers trapped in the QW have a tunneling time back to the charge centers in the barriers several orders of magnitude longer than the 82~MHz repetition rate of the laser used in time-resolved experiments. Therefore, the population of available acceptor states in the barrier is quickly fully depleted even when laser pulses with very weak photon density are used to excite the sample. Therefore, even under this pulsed excitation, the saturation value of the carrier population in the well is reached. We use excitation energy of 1.5072~eV in time-resolved experiments because at this photon energy only electrons from ionized acceptors can be excited to conduction band. Therefore, the background electron density is given by the concentration of ionized acceptors in the barrier, which is determined by growth conditions of he sample and is of the order of $10^{10}$cm$^{-2}$.

In order to study the dynamics of the luminescence in different density domains, we performed time-resolved experiments with a variety of absorbed photon densities
($10^8-10^{10}$~cm$^2$). Then,  to obtain the time evolution of the exciton, trion and plasma transition luminescence, for each time delay we spectrally integrate each transition. For that, weneed to resolve the exciton and trion overlapping lines. We obtain the pure exciton line thanks to the cw experiments (Fig.~1) with excitation at 1.5174~eV. Then, to attain the pure trion transition we substract the exciton line from photoluminescence sectrum obtained when the sample is excited at 1.5072~eV. In that way, we have the spectral limit to carry out the integration of pure exciton and trion trasitions at each time delay. In Fig.~2, we sho the time-evolution of the exciton, trion and plasma intensities obtained from this analysis.

The luminescence dynamics is governed by the temporal evolution of the population of free carriers, excitons and trions. We first make an inventory of all channels that couple those populations. Apart from  the known bi-molecular reaction $e+h\leftrightarrow X$, we identify two bi-molecular reactions ($X+e\leftrightarrow X^-$, $X+h\leftrightarrow X^+$) and two tri-molecular reactions ($2e+h\leftrightarrow X^-$, $e+2h\leftrightarrow X^+$) involving trions. The kinetics of these reactions is given in terms of five coupled rate equations:
\begin{align*}
 & \frac{d\mathbf{n}}{dt}=  - B \mathbf{n}\mathbf{p} -\frac{\mathbf{n}}{\tau_{\text{nr}}} -
  F^{X} - F_2^{X^{-}}  - F_3^{X^{-}}
    - F_3^{X^{+}}+ \frac{\mathbf{X}^{-}}{\tau_{\text{X}^-}}
  \\
 & \frac{d\mathbf{p}}{dt}= - B \mathbf{n}\mathbf{p} -\frac{\mathbf{p}}{\tau_{\text{nr}}}   -
  F^{X} - F_2^{X^{+}}- F_3^{X^{+}} - F_3^{X^{-}}
  + \frac{\mathbf{X}^{+}}{\tau_{\text{X}^+}}
  \\
 & \frac{d \mathbf{X}}{dt}= F^{X} - \frac{\mathbf{X}}{\tau_{\text{D}}}
  - F_2^{X^{-}} - F_2^{X^{+}}
  \\
 & \frac{d\mathbf{X^{-}}}{dt}= F_2^{X^{-}}+F_3^{X^{-}}
   - \frac{\mathbf{X}^{-}}{\tau_{\text{X}^-}}
  \\
 & \frac{d\mathbf{X^{+}}}{dt}=  F_2^{X^{+}} +  F_3^{X^{+}}
   - \frac{\mathbf{X}^{+}}{\tau_{\text{X}^+}},
\end{align*}
where the free carrier concentrations $\mathbf{n}$ (electrons) and $\mathbf{p}$ (holes) decay through --- in order of appearance --- the radiative and non-radiative recombination rates, the exciton formation rate $F^{X}$ and the trion formation rates $F_2^{X^{\alpha}}$ (bi-molecular) and $F_3^{X^{\alpha}}$ (tri-molecular), ${\alpha}=\{-,+$\}. The term $\mathbf{X}^{\alpha}/\tau_{\text{X}^{\alpha}}$ corresponds to the carriers recycled after the radiative decay of the trions. The exciton population $\mathbf{X}$ decays through radiative recombination and trion formation, while trion concentrations $\mathbf{X}^-$ and $\mathbf{X}^+$ decay through radiative recombination. The formation rates $F_X$, $F_2^{X^\alpha}$, $F_3^{X^\alpha}$ read
\begin{align*}\label{ratioeqnNXT}
F^{X}&=\gamma C \mathbf{n}\mathbf{p}- \gamma C K_X \mathbf{X},\\
F_2^{X^{-}} &=  A_2^- \mathbf{X}\mathbf{n} - A_2^- K_2^{-} \mathbf{X^{-}},\\
F_3^{X^{-}} &=  A_3^- \mathbf{n}\mathbf{n}\mathbf{p} - A_3^- K_3^{-} \mathbf{X^{-}},\\ F_2^{X^{+}} &=  A_2^+ \mathbf{X}\mathbf{p} - A_2^+ K_2^{+} \mathbf{X^{+}},\\
F_3^{X^{+}} &= A_3^+ \mathbf{n}\mathbf{p}\mathbf{p} - A_3^+ K_3^{+} \mathbf{X^{+}},
\end{align*}
where the equilibrium coefficients  $K_X$, $K_2^{\alpha}$, $K_3^{\alpha}$ have been introduced to ensure the steady-state solution of the rate equations \cite{Szczytko2004,Philips1996}.

\begin{figure}
  \includegraphics[width=\columnwidth]{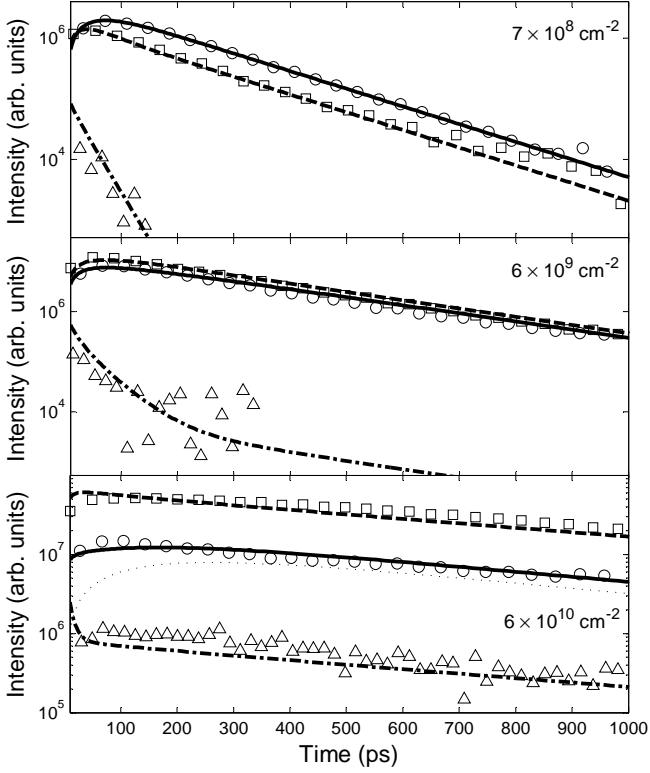}
\caption{\label{fig2} The intensity of the luminescence of trions (solid line, circles), excitons (dashed line, squares) and
plasma (dashed-dotted line, triangles) calculated according to rate
equations (lines) compared with experimental data (symbols) for $7\times10^{8}$,
$6\times10^{9}$, $6\times10^{10}$ cm$^{-2}$ absorbed photon densities. At $6\times10^{10}$ cm$^{-2}$ density, the best fit for trions after artificially enforcing $A_3=0$ (thin dashed line). $\mathbf{N}=10^{10}$ cm$^{-2}$.}
\end{figure}

A few considerations allow us to reduce significantly the number of free parameters.
The equilibrium coefficients can be calculated from a mass action law, exploiting the fact  that the trion binding energies for both ${X^{-}}$ and ${X^{+}}$ are equal~\cite{twotrions}. The values of the bi-molecular plasma recombination rate $B$ and bi-molecular exciton formation rate $\gamma C$ are known~\cite{Szczytko2004}. Due to the high quality of the sample we assume very long non-radiative decay time $\tau_{\text{nr}}$. Knowing the number of photons $\mathbf{N}_{h\nu}$ absorbed in our sample, we use the initial parameters $\mathbf{p}=\mathbf{N}_{h\nu}$ and $\mathbf{n}=\mathbf{N}_{h\nu}+\mathbf{N}$, where the excess electron concentration is estimated from impurity concentration: $\mathbf{N}=10^{10}$~cm$^{-2}$. We have introduced the equations of formation for positive trions for sake of completeness. However, our present measurements are not sensitive to the $X^+$ population and we have decided to equate $A_2^+$ with $A_2^-$ and $A_3^+$ with $A_3^-$. Once we have found an expression for the thermalized exciton and trion radiative decay times $\tau_D$ and $\tau_{X^\alpha}$,  $A_2$ and $A_3$ will be the only parameters left.  

We assume that excitons, trions and free carriers are thermalized and do share a same temperature $T_c$, different from the lattice temperature $T_l$. In our time resolved experiment, we use three  electron-hole pair densities $7\times 10^{8}$, $6\times 10^{9}$, $6\times 10^{10}$ cm$^{-2}$. At the highest density, $T_c$ is given by the exponential fit to the high energy tail of the free carrier luminescence. Tracing $T_c$ over 1000~ps returns 35~K in average. At lower densities, the rapid plasma relaxation prevents us from measuring $T_c$ that way. Yet, it can be trivially calculated if we depict the accumulated excess carriers trapped in the QW as a cold sea of electrons at $T_l$, in which the electron-hole pairs injected by the optical pump efficiently thermalize. We get 9~K and 16~K. At last, the temperature dependence of the radiative decay rates $\tau_{\text{D}}(T_c)$ and $\tau_{\text{X}^{\alpha}}(T_c)$ are most accurately described by the linear fits $\tau_{\text{D}}(T_c)=20\times T_c$~[ps] and $\tau_{\text{X}^{\alpha}}=78+7\times T_c$~[ps] (with $T_c$ in [K]). Apart from a factor 1.5 attributed to the Brag mirrors that enhances the coupling of the excitons and charged excitons to the field, and hence diminishes the radiative decay time, the agreement with the expected theoretical behavior is very good~\cite{Axel2000}. 

\begin{figure}
  \includegraphics[width=\columnwidth]{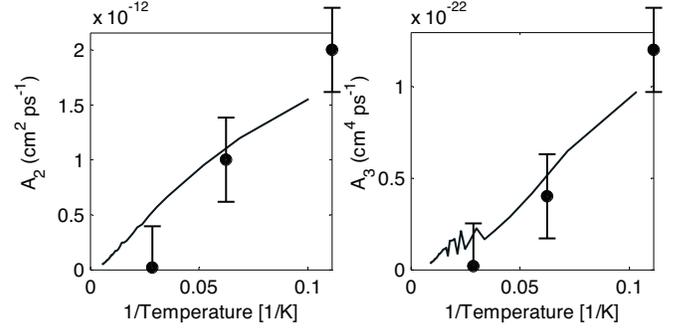}
  \caption{\label{fig3} The bi- and tri- molecular trion formation coefficients $A_2^{\alpha}$ and $A_3^{\alpha}$ as a function of the inverse carrier temperature. The dots correspond to the formation coefficients obtained from the fit. The line show the value expected from the theory. }
\end{figure}

We present the complete results of our calculations of excitonic,
trion and plasma luminescence intensity dynamics in
Fig.~\ref{fig2}. The excitonic luminescence intensity given by
$\mathbf{X}/\tau_{\text{D}}$, is denoted by a dashed line. The
luminescence intensities of  ${X^{-}}$ and ${X^{+}}$ are summed up ($X^{-}/\tau_{\text{X}^-}+X^{+}/\tau_{\text{X}^+}$) and denoted by solid lines. The free carrier luminescence $B\mathbf{np}$ is denoted by a dash-dotted line. Even with the necessary simplifications mentioned above, our rate equations provide a very good description of the observed time-resolved luminescence spectra. For instance for the smallest excitation densities the strongest transition comes from the trions, while for the larger densities the exciton transition dominates. Interestingly, for the lowest excitation
density, trion and exciton dynamics are much faster than for the highest, in striking contrast with the exciton dynamics in undoped QWs~\cite{Szczytko2004}. We may understand this fast luminescence dynamics in the low density regime: this is mostly the effect of the change of the temperature of carriers (and therefore of trions and excitons). If we impose the same temperature for all excitation densities, $\tau_{\text{D}}$ and $\tau_{\text{X}^{\alpha}}$ then stay constant over the densities, and we get, as expected, slower dynamics for lower than for higher densities.

We report in Fig.~\ref{fig3} the parameters $A_2$ and $A_3$ derived from our fit (dots). A theoretical calculation of the bi- and tri-molecular formation dependence on temperature was performed \cite{Berney2007} base on the formalism developed by Piermarocchi \emph{et al.} in the framework of exciton formation \cite{Piermarocchi1997}. The results of the calculation are shown on Fig.~\ref{fig3} (black lines) after having been multiplied by a factor 1.5. Experimental and theoretical results have the same temperature dependence, which reminds the behavior of the bi-molecular
recombination rate $B(T_c)$ . This directely comes from the momentum conservation of carriers which plays a very important role in both bi- and tri-molecular trion formation processes.  Secondly, it turns out that both bi- and tri-molecular channels are essential to the generation of trions. At high carrier densities, i.e. short times or large densities, the tri-molecular process even dominates the bimolecular. This is evidenced by the thin-dashed fit in Fig.~\ref{fig2}, calculated after having artificially eliminated the $A_3$ component. Finally, the trion formation time from an initial resonantly excited gas of exciton was measured in CdTe QWs for both $X^-$ \cite{Portella2003} and $X^+$ \cite{Kossacki2004} and turned out to be identical, which comforts our assumption $A_2^-=A_2^+$. The bimolecular coefficient drawn from those experiments ($A_2\approx 3\times 10^{-12}$~$\textrm{cm}^2/\textrm{ps}$) matches ours.

\begin{figure}
  \includegraphics[width=\columnwidth]{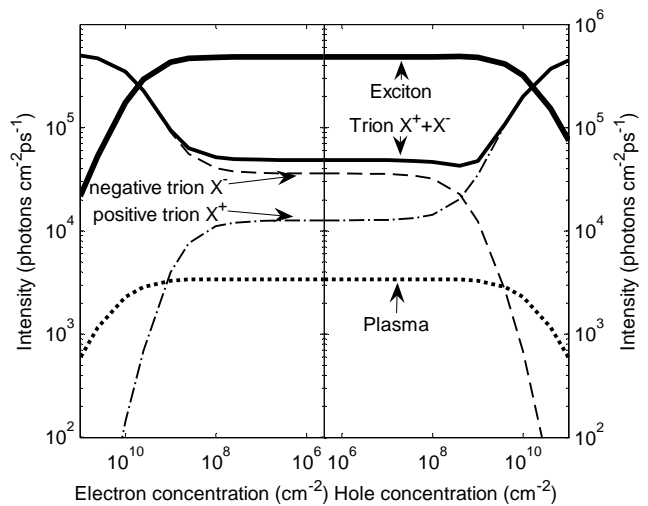}
\caption{\label{fig4} Exciton, trion
(${X^{-}}$ and ${X^{+}}$) and plasma cw-luminescence intensities calculated as a function
of the residual electron (left panel) and hole (right panel)
concentrations at $T_c$=9.0~K. A $5\times10^5$~photons/(cm$^2$ps) density was assumed~ \cite{przypis}.}
\end{figure}

We demonstrate the robustness of our model by some convincing predictions. In Fig.~\ref{fig4}, we apply our rate
equations model to the cw-luminescence of excitons, trions and free
carriers in an InGaAs QW. In the absence of excess carriers ($\mathbf{n}=\mathbf{p}$), the trion luminescence is about 20 times weaker than the exciton one,
which is in very good agreement with experiments~\cite{twotrions}. The difference in intensities of ${X^{-}}$ and ${X^{+}}$ cw-luminescence shown in Fig.~\ref{fig4} are the consequence of the difference between $K_2^{\alpha}$ and $K_3^{\alpha}$, imposed by the mass difference between positive and negative trions. The critical carrier concentration of about $10^{10}$ cm$^{-2}$ at which trions start to dominate the luminescence spectrum, does correspond to many experiments \cite{Kossacki2003}. What is new however, is that the crossing of $X^-$ and $X^+$ intensities, which has been observed experimentally \cite{twotrions}, does not occur at zero excess carrier density but is shifted toward some positive carrier density.

Applied to a system of excitons under resonant excitation, our model explains a very puzzling decay time of excitons, which is raised from 20~ps to 100~ps in the presence of an excess electron gas~\cite{Finkelstein1998}. Excitons and trions actually come into thermal equilibrium and decay together, which considerably stretches the decay time. 

In summary, we have shown that both bi- and tri-molecular processes are necessary to describe trion formation. We could quantify both formation rates from the experiment and show that their match the theoretical temperature dependance. We obtained new insight on ${X^{-}}$ and ${X^{+}}$ luminescence intensities at low excess carrier densities. The model turned out to be also perfectly applicable to other experiments.

Acknowledgments: we acknowledge financial support from FNRS within
quantum photonics NCCR. We thank D.~Y.~Oberli and M.~Richard for enlightening discussions.

\end{document}